\documentclass[%
reprint,
 amsmath,amssymb,
 jcp,
 aip,
 floatfix,
 showkeys,
]{revtex4-1}

\usepackage{graphicx}
\usepackage{dcolumn}
\usepackage{bm}
\usepackage{bbold}
\usepackage{bbm}
\usepackage{hyperref}
\hypersetup{colorlinks = true,allcolors = {blue}}

\begin{document}
\title{Communication: Truncated Non-Bonded Potentials Can Yield Unphysical \texorpdfstring{\\}{} Behavior in Molecular Dynamics Simulations of Interfaces}

\author{Martin Fitzner}
\affiliation{Thomas Young Centre, London Centre for Nanotechnology and Department of Physics and Astronomy, University College London, Gower Street London WC1E 6BT, United Kingdom}

\author{Laurent Joly}
\affiliation{Univ Lyon, Universit\'{e} Claude Bernard Lyon 1, CNRS, Institut Lumi\`{e}re Mati\`{e}re, F-69622 Villeurbanne, France}

\author{Ming Ma}
\affiliation{Department of Mechanical Engineering, State Key Laboratory of Tribology and Center for Nano and Micro Mechanics, Tsinghua University, Beijing 100084, China}

\author{Gabriele C. Sosso}
\affiliation{Department of Chemistry and Centre for Scientific Computing, University of Warwick, Gibbet Hill Road, Coventry CV4 7AL, United Kingdom}

\author{Andrea Zen}
\affiliation{Thomas Young Centre, London Centre for Nanotechnology and Department of Physics and Astronomy, University College London, Gower Street London WC1E 6BT, United Kingdom}

\author{Angelos Michaelides}
\email{angelos.michaelides@ucl.ac.uk}
\affiliation{Thomas Young Centre, London Centre for Nanotechnology and Department of Physics and Astronomy, University College London, Gower Street London WC1E 6BT, United Kingdom}

\date{\today}

\begin{abstract}
Non-bonded potentials are included in most force fields and therefore widely used in classical molecular dynamics (MD) simulations of materials and interfacial phenomena. It is commonplace to truncate these potentials for computational efficiency based on the assumption that errors are negligible for reasonable cutoffs or compensated for by adjusting other interaction parameters. Arising from a metadynamics study of the wetting transition of water on a solid substrate we find that the influence of the cutoff is unexpectedly strong and can change the character of the wetting transition from continuous to first order by creating artificial metastable wetting states. Common cutoff corrections such as the use of a force switching function, a shifted potential or a shifted force do not avoid this. Such a qualitative difference urges caution and suggests that using truncated non-bonded potentials can induce unphysical behavior that cannot be fully accounted for by adjusting other interaction parameters.     
\end{abstract}

\keywords{Molecular Dynamics, Interfaces, Force Fields, Free Energy}
\maketitle
Short- to medium-range potentials such as the Lennard-Jones~\cite{jones1924determination} or the Buckingham~\cite{buckingham1938classical} potential are the backbone of classical MD simulations. They represent Pauli repulsion as well as non-directional dispersion attraction and there exist multiple flavors implemented in most MD codes under the term of non-bonded interactions. In practice there is a need to truncate these potentials since the number of neighbors that have to be considered for each entity grows enormously, drastically increasing the computational cost for the force calculation. Truncating between $r_\mathrm{c}$ = 2.5$\sigma$ and 3.5$\sigma$, where $\sigma$ is the characteristic interaction range, is a very common practice in MD studies~\cite{frenkel2001understanding} and has become the minimum standard, assuming that errors arising from this are small enough. Several studies have reported that with these settings significant problems can arise. For instance the truncation can alter the phase diagram of the Lennard-Jones system~\cite{smit1992phase,wang2008homogeneous} or yield different values for interfacial free energies~\cite{ismail2007surface,valeriani2007comparison,ghoufi2016computer,ghoufi2017importance,marcello2017LongRange}. These effects are quantitative in nature, meaning that they can in certain circumstances be analytically corrected for~\cite{sun1998compass,sinha2003surface,werth2015long} or compensated for by other interaction parameters such as interaction strength or interaction range. The latter is important for the development of force fields where non-bonded potentials are often included and the cutoff can be seen as another fitting parameter. Naturally, a parametrization with a small cutoff would be preferred to another one if they deliver equal accuracy. This however is only true in the assumption that the underlying physical characteristics that are created by truncated and longer ranging potentials are the same.

In this work we investigated the influence of the cutoff for the interfacial phenomenon of water-wetting on a solid substrate. We found that the effect of the cutoff of the water-substrate interaction was not only unexpectedly strong, but also changed the fundamental physics of the wetting transition in an unprecedented way by creating metastable wetting states that have also never been seen in experiments. We show that proposed cutoff corrections such as the use of a force switching function, a shifted potential or a shifted force did not fix this and could even worsen the effect. This finding shows that atomistic simulations of interfaces need to be treated with great care since unphysical behavior could occur and easily remain undetected. This is particularly relevant since a large number of MD studies using truncated potentials are reported each year. Our results suggest the use of much larger-than-common cutoffs or long-range versions of non-bonded potentials in MD studies of wetting and interfacial phenomena. 

\begin{figure}[!t]
	\centerline{\includegraphics[width=6.6cm]{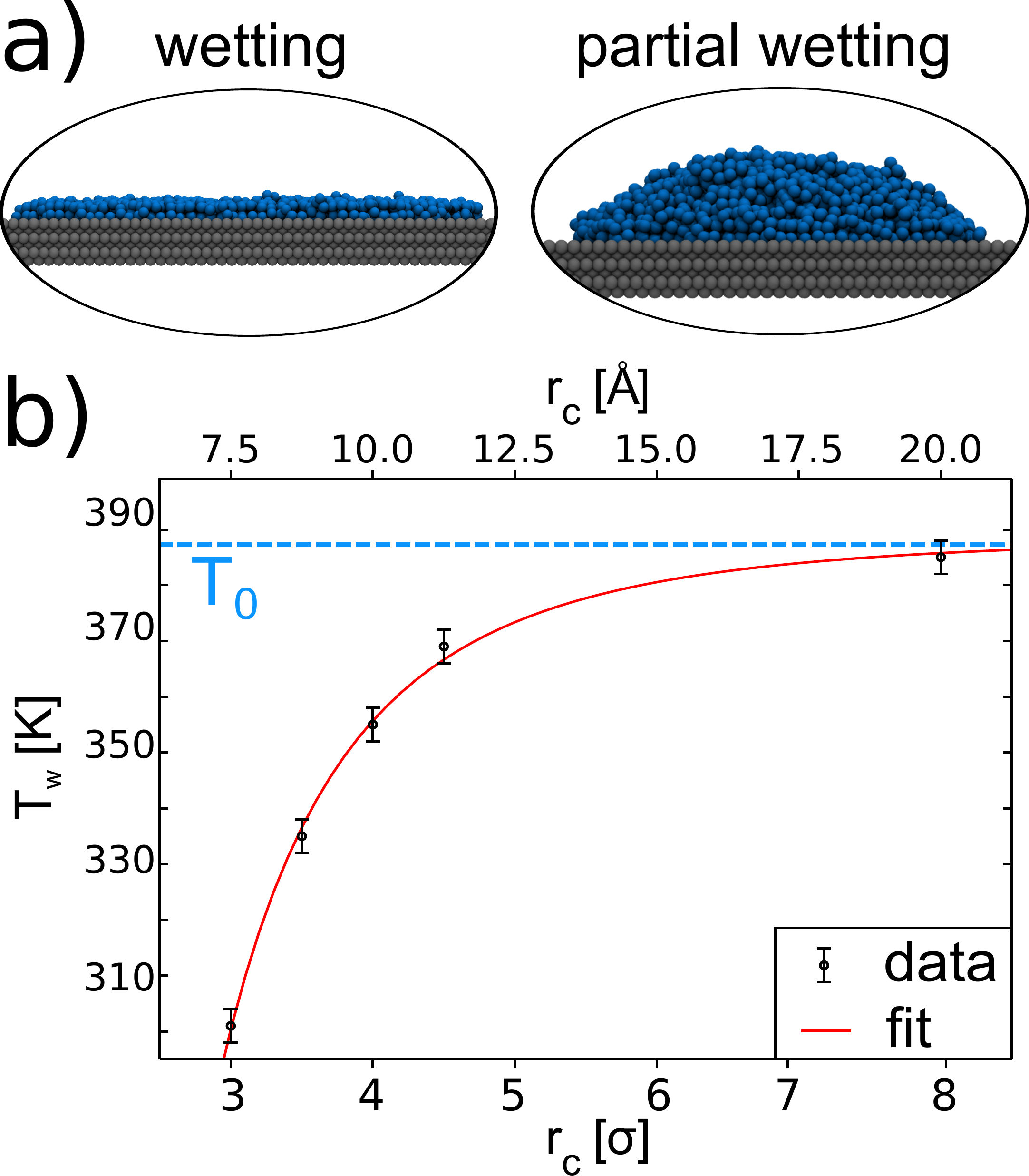}}
	\caption{a) Side view of the two wetting states for the small droplet. Water is blue and surface atoms are gray. b) Temperature of the wetting transition $T_\mathrm{w}$ (points) versus cutoff radius $r_\mathrm{c}$ and fit (red line). The $T_\mathrm{w}$ were obtained from the free energy profiles (see text) and we estimate errors to be $\pm 3$~K. $T_0$ is the converged wetting temperature.}
	\label{FIG_1_TRANSITION_TEMPERATURE}
\end{figure}
\begin{figure*}[!t]	\centerline{\includegraphics[width=16.2cm]{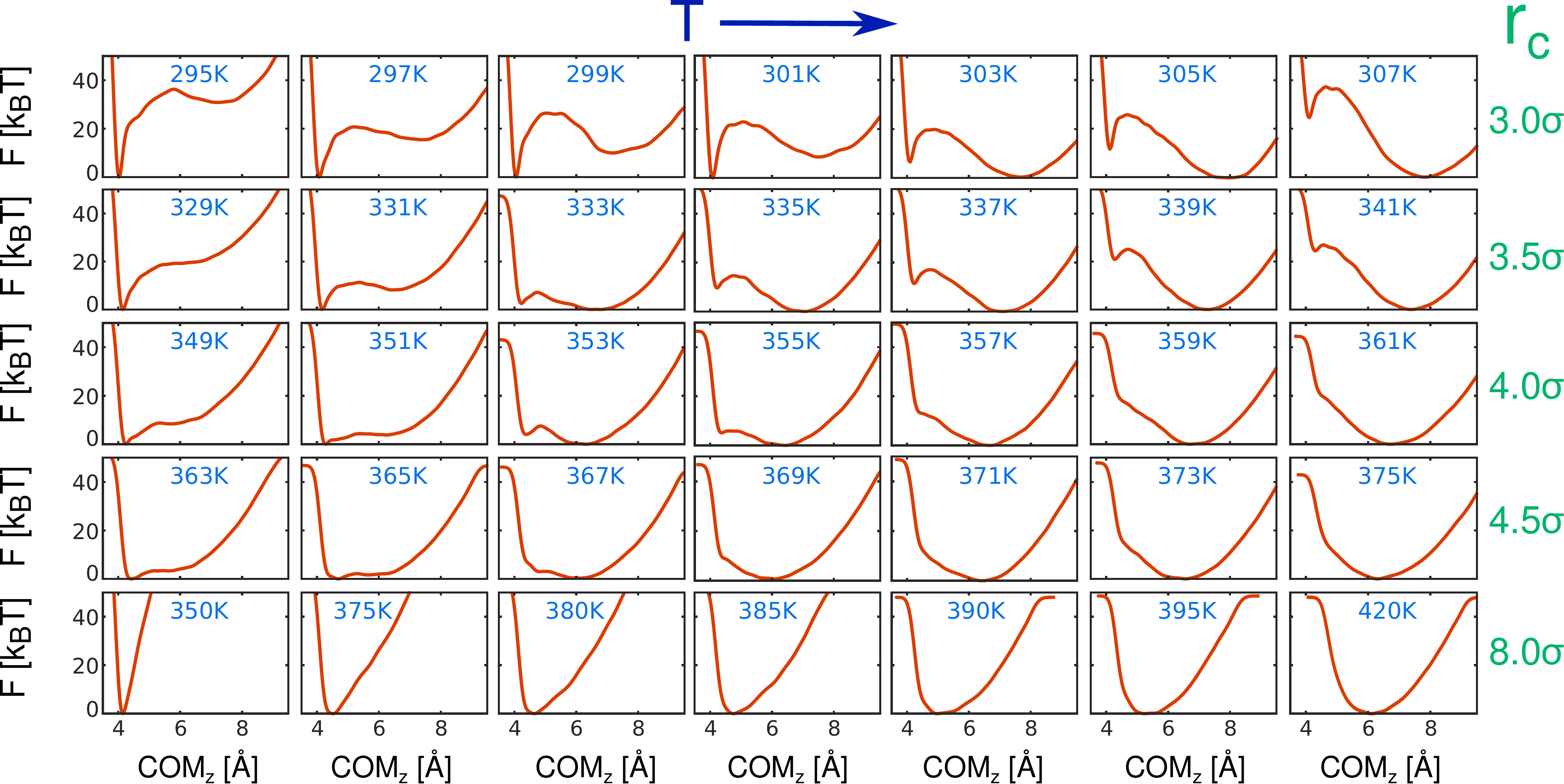}}
	\caption{Free energy profiles of wetting for different cutoffs in a small temperature range around the respective transition temperature $T_\mathrm{w}$ (generally at or near the central column for each system). As collective variable we chose the center of mass of the water droplet ($\mathrm{COM}_\mathrm{z}$, substrate at $z = 0$). We note that for the largest cutoff of $8\sigma$ the temperature range is slightly larger to highlight the shape of the free energy profile for complete and partial wetting.}
	\label{FIG_2_FREEENERGY}
\end{figure*}
We investigated two droplets comprised of 3000 and 18000 water molecules which were represented by the coarse-grained mW model~\cite{molinero_water_2009}, on top of a rigid, pristine fcc(100) surface (lattice parameter 4.15~\r{A}). Whilst this substrate does not aim at representing any particular material, similar systems have been used to study ice nucleation~\cite{reinhardt2012free,cox2015molecular1,cox2015molecular2,fitzner_many_2015} or water-metal interfaces~\cite{heinz2008accurate,xu2015nanoscale}. The simulation cell had dimensions $17\times 17 \times 11$~nm which is enough to avoid interaction of the water molecules with their periodic images for all wetting states. Even though the liquid is rather non-volatile even at the highest temperature considered, we employed a reflective wall at the top of the cell to avoid evaporation and mimic experimental conditions. Our simulations were performed with the LAMMPS code~\cite{plimpton1995fast}, integrating the equations of motion with a timestep of 10~fs. This rather large timestep is commonly used in combination with the mW model and is acceptable for our system since during NVE simulations the total energy drift was found to be only about $2\times10^{-9}$~eV per water molecule per ps. In addition, we verified that we obtain the same results using standard protocols for updating the neighbor lists compared with unconditionally updating them every timestep. All production simulations were performed in the NVT ensemble with constant temperature maintained by a ten-fold Nos\'{e}-Hoover chain~\cite{martyna1992nose} with a relaxation time of 1~ps. The substrate-water interaction was given by a distance ($r$) dependent Lennard-Jones potential
\begin{equation}
	U_\mathrm{LJ}(r)=4\epsilon\left[ \left(\frac{\sigma}{r}\right)^{12} - \left(\frac{\sigma}{r}\right)^{6} \right] \end{equation}
with $\epsilon = $ 29.5~meV, $\sigma$ = 2.5~\r{A} truncated at a cutoff $r_\mathrm{c}$. This resulted in a maximum interaction energy of 154~meV for an adsorbed water monomer (weakly depending on the cutoff). Additionally we performed well-tempered metadynamics simulations~\cite{laio2002escaping,barducci2008well} for the smaller droplet with the PLUMED2 code~\cite{tribello2014plumed}. In these simulations the Gaussian height, width, bias-factor and deposition stride were $2.16$~meV, $0.15$~\r{A}, 20 and $20$~ps respectively. Metadynamics is usually applied to drive rare events such as nucleation~\cite{sosso2016crystal,sosso2016ice,tribello2017analyzing,cheng2017bridging} or protein folding~\cite{bussi2006free,laio2008metadynamics}. In our systems, this method helped to uncover the underlying free energy profile of wetting.

We studied the wetting behavior of the larger droplet by performing standard MD runs at different temperatures first. As starting configurations we chose either a flat water film in direct contact or a spherical droplet placed above the substrate. Within at most 5~ns the simulation was equilibrated and a seemingly stable configuration was reached, where the water is either wetting (contact angle $\theta = 0^{\circ}$) or partially wetting ($0^\circ < \theta < 180^\circ$). An illustration of the two wetting states can be found in figure~\ref{FIG_1_TRANSITION_TEMPERATURE}a. Initially we employed a radial cutoff at $r_\mathrm{c} = 3.0\sigma$ for the water-substrate interaction. With this setting we found that interestingly a wetting transition happened at finite angle $\theta_0 \approx 23^\circ$, i.e. a smaller non-zero contact angle was not possible. This behavior cannot be explained by the standard Young's equation.

However, upon increasing the cutoff we found that the wetting behavior drastically changed. First, the wetting temperature $T_\mathrm{w}$ at which the wetting transition took place increased as we increased the cutoff (figure~\ref{FIG_1_TRANSITION_TEMPERATURE}b). Whilst $T_\mathrm{w}$ shows a clear convergence behavior with $r_\mathrm{c}$, it is unexpectedly slow. A reasonably converged wetting temperature $T_0$ is only reached for $r_\mathrm{c} > 7\sigma$. Second, we noticed that for an increasing cutoff the minimum possible contact angle $\theta_0$ got smaller and eventually vanished. Most importantly, we also found that for temperatures around $T_\mathrm{w}$ the stable configuration that was reached after the 5~ns could depend on the starting configuration for smaller cutoffs, while for larger $r_\mathrm{c}$ it always reached the same state. This suggests that for small $r_\mathrm{c}$ we actually found metastable wetting states that are absent for large $r_\mathrm{c}$. This also means that $T_\mathrm{w}$ cannot naively be defined through visual analysis of trajectories at different temperatures but needs to be defined by the free energy of wetting. For a first order phase transition we define $T_\mathrm{w}$ to be the temperature where the two basins (corresponding to wetting and partial wetting) have the same free energy. For a continuous phase transition $T_\mathrm{w}$ is the temperature where the single basin represents a contact angle of $\theta = 0^\circ$ for $T < T_\mathrm{w}$ and $\theta > 0^\circ$ for $T > T_\mathrm{w}$.

Understanding the character of these wetting states with standard MD can prove difficult as the dependence on the starting configuration always leaves doubt on the outcome of the equilibrated configuration obtained from it. To clarify, we show the results from the metadynamics simulations in figure~\ref{FIG_2_FREEENERGY}. As a collective variable we chose the z-component of the center of mass of the water droplet ($\mathrm{COM}_\mathrm{z}$), where z is the surface normal direction. While this choice is not equivalent to the contact angle (as they are related in a non-linear manner) it is clear that significantly different values for $\mathrm{COM}_\mathrm{z}$ correspond to different contact angles and can therefore distinguish the different wetting states. For the smallest cutoff at $T_\mathrm{w}$ and around we found that two basins coexist, one being the flat film (COM$_\mathrm{z} \approx$ 4~\r{A}) and the other being a droplet with certain contact angle (COM$_\mathrm{z} \gtrsim 5$~\r{A}). These two states are separated by a significant barrier larger than 20~$k_\mathrm{B}T$, which explains why we observed metastable states in the unbiased simulations for small $r_\mathrm{c}$. This corresponds to a first-order phase transition between the wetting states. The occurrence of a minimum possible contact angle $\theta_0$ is explained by the existence of the second basin, which does not approach the wetting basin, but rather becomes less stable as temperature changes. However, this character faded as we increased $r_\mathrm{c}$. The barrier became smaller and the distance between the basins got smaller. For the largest cutoff investigated ($8\sigma$) we clearly see that only a single basin exists that changes its position with temperature. As a result no metastable wetting states exist and the phase transition is continuous. We note that in this case the estimate of $T_\mathrm{w}$ is more difficult than for the first order transitions, however in this work we aim at presenting qualitative results and from figure~\ref{FIG_2_FREEENERGY} it is clear that $T_\mathrm{w}$ is higher than for the smaller cutoffs.

Only the results for the largest cutoff are in agreement with the fact that water wetting transitions are generally continuous when probed in experiments~\cite{bonn2001wetting,friedman2013wetting} and finite-angle wetting transitions have, to the best of our knowledge, never been observed experimentally. Therefore, the correct qualitative wetting behavior in our system is not achieved with standard cutoffs and if undetected could potentially lead to false conclusions. Differences between short and long-ranged interactions have been highlighted for other interfacial phenomena, such as drying~\cite{evans2016critical} or grain boundary melting~\cite{caupin2008absence}.

\begin{figure}[!t]
	\centerline{\includegraphics[height=6.8cm]{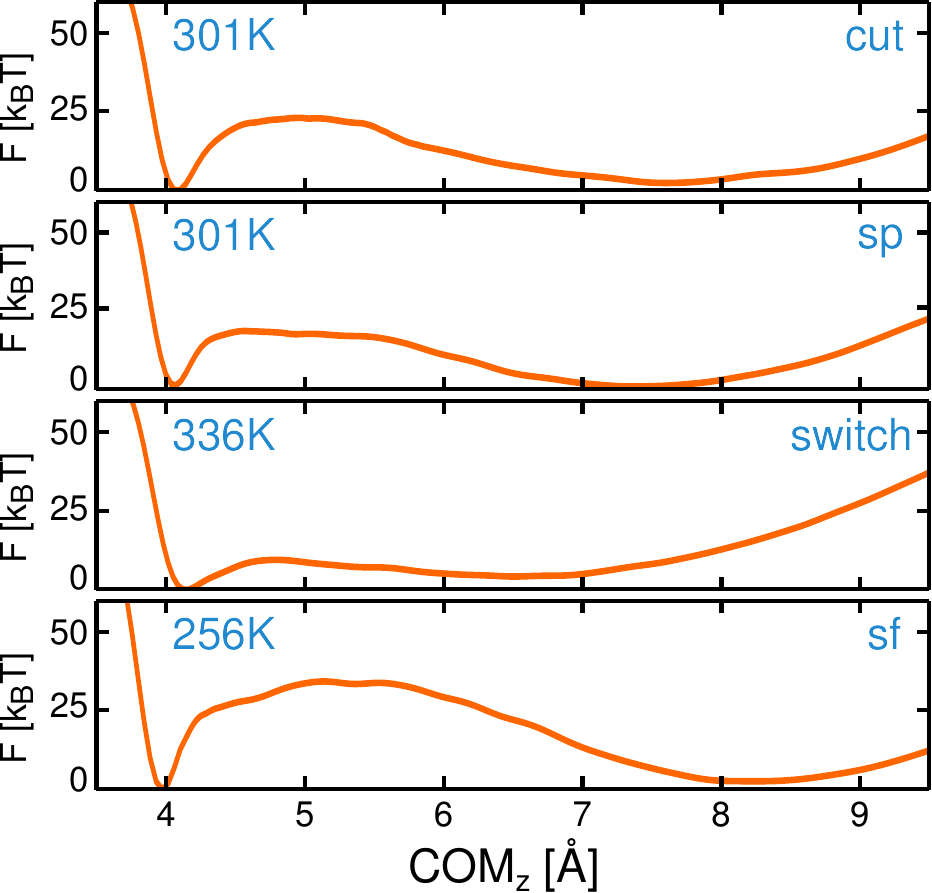}}
	\caption{Free energy profiles of wetting approximately at the transition temperature with uncorrected setup (cut) and for different correction schemes [shifted potential (sp), force switch (switch) and shifted force (sf)] applied with a cutoff at $3\sigma$. None of the schemes show the correct behavior, which is shown in figure~\ref{FIG_2_FREEENERGY} to be a single basin.}
	\label{FIG_3_SHIFTED}
\end{figure}

We further study the effect of the most commonly used correction schemes to cutoffs: 
\begin{enumerate}
\item A shifted potential (sp) which ensures that the value of the potential energy $U$ does not jump at the cutoff distance, given by:
\begin{align}
	U_\mathrm{sp}(r) &= U_\mathrm{LJ}(r) - U_\mathrm{LJ}(r_\mathrm{c})
\end{align}
The corresponding force $F$ remains unaltered:
\begin{align}
	F_\mathrm{sp}(r) &= F_\mathrm{LJ}(r)						
\end{align}
\item A switching function (switch) which brings the force to zero between an inner $r_\mathrm{c,1}$ and an outer cutoff $r_\mathrm{c,2}$ (we chose $3\sigma$ and $4\sigma$):
\begin{align}
	F_\mathrm{switch}(r) &= F_\mathrm{LJ}(r)					&r \le r_\mathrm{c,1}&\\
	F_\mathrm{switch}(r) &= \sum_{k=0}^3 C_k(r-r_\mathrm{c,1})^{k} &r_\mathrm{c,1} < r \le r_\mathrm{c,2}& \nonumber
\end{align}
where $C_k$ are constants determined to ensure a smooth behavior~\cite{plimpton1995fast}.
\item A shifted-force potential (sf), which ensures that force and potential do not jump:
\begin{align}
	U_\mathrm{sf}(r) &= U_\mathrm{LJ}(r) - U_\mathrm{LJ}(r_\mathrm{c}) - (r - r_\mathrm{c})F_\mathrm{LJ}(r_\mathrm{c}) 	\\
	F_\mathrm{sf}(r) &= F_\mathrm{LJ}(r) - F_\mathrm{LJ}(r_\mathrm{c})	\nonumber
\end{align}
\end{enumerate}
The latter approach was found to give good results for a homogeneous system and even allowed for a reduction of the cutoff~\cite{toxvaerd2011communication}. Our results for these three corrections can be found in figure~\ref{FIG_3_SHIFTED}. By definition and thus unsurprisingly, the shifted potential does not yield any significant difference (where the remaining minor deviations are due to the metadynamics sampling) over the plain cutoff since forces remain unaltered. The smooth cutoff via switching function seems to improve the situation, however the fact that the transition temperature lies between the ones we found for a plain cutoff at $3\sigma$ and $4\sigma$ suggests that the improvement stems from the effectively increased interaction range rather than the fact that the force vanishes smoothly. Interestingly, the shifted force with the same cutoff performs worst out of all candidates as the barrier increases by a factor of two, which increases the likelihood that simulations are performed in the metastable state without realizing it. The fact that none of the considered correction schemes significantly improved the character of the wetting free energy profile leads us to conclude that it is not the way in which the cutting is done that matters most, but rather the effective cutoff distance as well as the overall interaction strength at that distance. 

As an initial attempt to understand the results obtained we looked at the potential energies of the various systems with the different cutoffs considered. This, however, did not reveal any obvious explanation. One possible interpretation for the creation of metastable states in our systems with shorter cutoff can be obtained by considering the droplet state (not assuming anything about the stability relative to the film state). For a transition towards the film state, there needs to be thermal fluctuations of water molecules that are above the contact layer in the downwards direction (the fact that COM$_\mathrm{z}$ has proven a good reaction coordinate supports this statement). With an infinite interaction range all molecules that are loosing height contribute to these fluctuations since they have an interaction with the substrate. Therefore we expect the interaction energy to change monotonically and the free energy to follow monotonically either up or down depending on the balance of the interfacial free energies (see figure~\ref{FIG_2_FREEENERGY}, $r_\mathrm{c} = 8\sigma$). But if the interaction range is finite, not all molecules contribute to an increased interaction with the substrate even if they decrease their height (and subsequently weaken the water-water interaction of the system by leading to deviations from a perfect spherical droplet). In other words, there is a minimum distance from the substrate that has to be surpassed by a molecule for it to contribute to a fluctuation increasing the interaction energy, otherwise it will (on average) actually decrease the total interaction energy. This minimum fluctuation for a single molecule translates into the macroscopic states (droplet and film) being connected by a barrier shaped free energy profile rather than a monotonic one (see figure~\ref{FIG_2_FREEENERGY}, $r_\mathrm{c} = 3\sigma$). The entropic contributions to the free energy are unlikely to change this, since they are essentially dominated by the environment a molecule is in (quasi-static contact layer or quasi-liquid water on top). The entropic change between these two states will be monotonic for a single water molecule and therefore also for the whole droplet.

Finding a general recipe for how to avoid such unphysical wetting states is difficult. Other aspects like e.g. the substrate density or the liquid-liquid interaction strength will have an influence on how strongly the fluctuations in the droplet state are affected by $r_\mathrm{c}$. Generally, cutoffs that are deemed acceptable from the inter-molecular perspective do not necessarily mean that the interaction between macroscopic states such as a film/droplet and a substrate is sufficiently captured. This is especially important in an interfacial simulation setting such as a slab, where a cutoff-caused change in interaction from the substrate side is not compensated by an equal change from the vacuum side. Consequently, only employing much larger cutoffs or techniques to calculate the long-range part of the dispersion force~\cite{in2007application,isele2012development,isele2013reconsidering} can ensure that unphysical effects are avoided. A minimal sanity check for future wetting studies could be to start simulations from both a wetting film and a spherical liquid snapshot. If both of them end up in the same configuration the existence of an unphysical metastable wetting state is unlikely.

In light of the vast amount of work that is done in the MD community using similar interactions, our findings urge extreme caution when dealing with truncated non-bonded potentials in simulations of interfacial phenomena. We have seen both quantitative and qualitative differences for the wetting transition. The former could be accounted for by changing other interaction parameters to reproduce the transition at the right temperature $T_0$. This assumption is fundamental to fitting force fields with truncated potentials to obtain quantitative agreement with e.g. experimental values. But it does not hold for the character of the transition because it arises purely from the value of the cutoff itself. If the resulting metastability of states remains undetected, the use of truncated interaction potentials could lead to wrong inferences about physical properties being made. While this conclusion has resulted from a simulation of wetting, similar implications could hold for other interfacial phenomena such as capillary flow~\cite{joly2011capillary,gravelle2016anomalous}, evaporation/condensation~\cite{hens2014nanoscale,nagayama2015molecular}, mixtures~\cite{iyer2013computer,tran2014molecular,radu2017enhanced} or heterogeneous nucleation~\cite{reinhardt2014effects,cabriolu2015ice,bi2016heterogeneous,qiu2017strength,bourque2017heterogeneous} where it is commonplace to use truncated interactions.

\begin{acknowledgments}
This work was supported by the European Research Council under the European Union's Seventh Framework Programme (FP/2007-2013) / ERC Grant Agreement number 616121 (HeteroIce project). A.M. is supported by the  Royal Society through a Royal Society Wolfson Research Merit Award. We are grateful for computational resources provided by the London Centre for Nanotechnology and the Materials Chemistry Consortium through the EPSRC grant number EP/L000202. L.J. is supported by the French Ministry of Defense through the project DGA ERE number 2013.60.0013 and by the LABEX iMUST (ANR-10-LABX-0064) of Universit\'{e} de Lyon, within the program ``Investissements d'Avenir'' (ANR-11-IDEX-0007) operated by the French National Research Agency (ANR). M.M. is supported by the Thousand Young Talent Program from the Organization Department of the CPC Central Committee.
\end{acknowledgments}

\bibliography{main}

\begin{thebibliography}{51}%
\makeatletter
\providecommand \@ifxundefined [1]{%
 \@ifx{#1\undefined}
}%
\providecommand \@ifnum [1]{%
 \ifnum #1\expandafter \@firstoftwo
 \else \expandafter \@secondoftwo
 \fi
}%
\providecommand \@ifx [1]{%
 \ifx #1\expandafter \@firstoftwo
 \else \expandafter \@secondoftwo
 \fi
}%
\providecommand \natexlab [1]{#1}%
\providecommand \enquote  [1]{``#1''}%
\providecommand \bibnamefont  [1]{#1}%
\providecommand \bibfnamefont [1]{#1}%
\providecommand \citenamefont [1]{#1}%
\providecommand \href@noop [0]{\@secondoftwo}%
\providecommand \href [0]{\begingroup \@sanitize@url \@href}%
\providecommand \@href[1]{\@@startlink{#1}\@@href}%
\providecommand \@@href[1]{\endgroup#1\@@endlink}%
\providecommand \@sanitize@url [0]{\catcode `\\12\catcode `\$12\catcode
  `\&12\catcode `\#12\catcode `\^12\catcode `\_12\catcode `\%12\relax}%
\providecommand \@@startlink[1]{}%
\providecommand \@@endlink[0]{}%
\providecommand \url  [0]{\begingroup\@sanitize@url \@url }%
\providecommand \@url [1]{\endgroup\@href {#1}{\urlprefix }}%
\providecommand \urlprefix  [0]{URL }%
\providecommand \Eprint [0]{\href }%
\providecommand \doibase [0]{http://dx.doi.org/}%
\providecommand \selectlanguage [0]{\@gobble}%
\providecommand \bibinfo  [0]{\@secondoftwo}%
\providecommand \bibfield  [0]{\@secondoftwo}%
\providecommand \translation [1]{[#1]}%
\providecommand \BibitemOpen [0]{}%
\providecommand \bibitemStop [0]{}%
\providecommand \bibitemNoStop [0]{.\EOS\space}%
\providecommand \EOS [0]{\spacefactor3000\relax}%
\providecommand \BibitemShut  [1]{\csname bibitem#1\endcsname}%
\let\auto@bib@innerbib\@empty
\bibitem [{\citenamefont {Jones}(1924)}]{jones1924determination}%
  \BibitemOpen
  \bibfield  {author} {\bibinfo {author} {\bibfnamefont {J.~E.}\ \bibnamefont
  {Jones}},\ }in\ \href {\doibase 10.1098/rspa.1924.0082} {\emph {\bibinfo
  {booktitle} {Proc. R. Soc. A}}},\ Vol.\ \bibinfo {volume} {106}\ (\bibinfo
  {organization} {The Royal Society},\ \bibinfo {year} {1924})\ pp.\ \bibinfo
  {pages} {463--477}\BibitemShut {NoStop}%
\bibitem [{\citenamefont {Buckingham}(1938)}]{buckingham1938classical}%
  \BibitemOpen
  \bibfield  {author} {\bibinfo {author} {\bibfnamefont {R.~A.}\ \bibnamefont
  {Buckingham}},\ }in\ \href {\doibase 10.1098/rspa.1938.0173} {\emph {\bibinfo
  {booktitle} {Proc. R. Soc. A}}},\ Vol.\ \bibinfo {volume} {168}\ (\bibinfo
  {organization} {The Royal Society},\ \bibinfo {year} {1938})\ pp.\ \bibinfo
  {pages} {264--283}\BibitemShut {NoStop}%
\bibitem [{\citenamefont {Frenkel}\ and\ \citenamefont
  {Smit}(2001)}]{frenkel2001understanding}%
  \BibitemOpen
  \bibfield  {author} {\bibinfo {author} {\bibfnamefont {D.}~\bibnamefont
  {Frenkel}}\ and\ \bibinfo {author} {\bibfnamefont {B.}~\bibnamefont {Smit}},\
  }\href@noop {} {\emph {\bibinfo {title} {Understanding Molecular Simulation:
  From Algorithms to Applications}}},\ Vol.~\bibinfo {volume} {1}\ (\bibinfo
  {publisher} {Academic press},\ \bibinfo {year} {2001})\BibitemShut {NoStop}%
\bibitem [{\citenamefont {Smit}(1992)}]{smit1992phase}%
  \BibitemOpen
  \bibfield  {author} {\bibinfo {author} {\bibfnamefont {B.}~\bibnamefont
  {Smit}},\ }\href {\doibase 10.1063/1.462271} {\bibfield  {journal} {\bibinfo
  {journal} {J. Chem. Phys.}\ }\textbf {\bibinfo {volume} {96}},\ \bibinfo
  {pages} {8639} (\bibinfo {year} {1992})}\BibitemShut {NoStop}%
\bibitem [{\citenamefont {Wang}, \citenamefont {Valeriani},\ and\ \citenamefont
  {Frenkel}(2008)}]{wang2008homogeneous}%
  \BibitemOpen
  \bibfield  {author} {\bibinfo {author} {\bibfnamefont {Z.-J.}\ \bibnamefont
  {Wang}}, \bibinfo {author} {\bibfnamefont {C.}~\bibnamefont {Valeriani}}, \
  and\ \bibinfo {author} {\bibfnamefont {D.}~\bibnamefont {Frenkel}},\ }\href
  {\doibase 10.1021/jp807727p} {\bibfield  {journal} {\bibinfo  {journal} {J.
  Phys. Chem. B}\ }\textbf {\bibinfo {volume} {113}},\ \bibinfo {pages} {3776}
  (\bibinfo {year} {2008})}\BibitemShut {NoStop}%
\bibitem [{\citenamefont {Ismail}\ \emph {et~al.}(2007)\citenamefont {Ismail},
  \citenamefont {Tsige}, \citenamefont {Veld~In't},\ and\ \citenamefont
  {Grest}}]{ismail2007surface}%
  \BibitemOpen
  \bibfield  {author} {\bibinfo {author} {\bibfnamefont {A.~E.}\ \bibnamefont
  {Ismail}}, \bibinfo {author} {\bibfnamefont {M.}~\bibnamefont {Tsige}},
  \bibinfo {author} {\bibfnamefont {P.~J.}\ \bibnamefont {Veld~In't}}, \ and\
  \bibinfo {author} {\bibfnamefont {G.~S.}\ \bibnamefont {Grest}},\ }\href
  {\doibase 10.1080/00268970701779663} {\bibfield  {journal} {\bibinfo
  {journal} {Mol. Phys.}\ }\textbf {\bibinfo {volume} {105}},\ \bibinfo {pages}
  {3155} (\bibinfo {year} {2007})}\BibitemShut {NoStop}%
\bibitem [{\citenamefont {Valeriani}, \citenamefont {Wang},\ and\ \citenamefont
  {Frenkel}(2007)}]{valeriani2007comparison}%
  \BibitemOpen
  \bibfield  {author} {\bibinfo {author} {\bibfnamefont {C.}~\bibnamefont
  {Valeriani}}, \bibinfo {author} {\bibfnamefont {Z.-J.}\ \bibnamefont {Wang}},
  \ and\ \bibinfo {author} {\bibfnamefont {D.}~\bibnamefont {Frenkel}},\ }\href
  {\doibase 10.1080/08927020701579352} {\bibfield  {journal} {\bibinfo
  {journal} {Mol. Sim.}\ }\textbf {\bibinfo {volume} {33}},\ \bibinfo {pages}
  {1023} (\bibinfo {year} {2007})}\BibitemShut {NoStop}%
\bibitem [{\citenamefont {Ghoufi}, \citenamefont {Malfreyt},\ and\
  \citenamefont {Tildesley}(2016)}]{ghoufi2016computer}%
  \BibitemOpen
  \bibfield  {author} {\bibinfo {author} {\bibfnamefont {A.}~\bibnamefont
  {Ghoufi}}, \bibinfo {author} {\bibfnamefont {P.}~\bibnamefont {Malfreyt}}, \
  and\ \bibinfo {author} {\bibfnamefont {D.~J.}\ \bibnamefont {Tildesley}},\
  }\href {\doibase 10.1039/C5CS00736D} {\bibfield  {journal} {\bibinfo
  {journal} {Chem. Soc. Rev.}\ }\textbf {\bibinfo {volume} {45}},\ \bibinfo
  {pages} {1387} (\bibinfo {year} {2016})}\BibitemShut {NoStop}%
\bibitem [{\citenamefont {Ghoufi}\ and\ \citenamefont
  {Malfreyt}(2017)}]{ghoufi2017importance}%
  \BibitemOpen
  \bibfield  {author} {\bibinfo {author} {\bibfnamefont {A.}~\bibnamefont
  {Ghoufi}}\ and\ \bibinfo {author} {\bibfnamefont {P.}~\bibnamefont
  {Malfreyt}},\ }\href {\doibase 10.1063/1.4976964} {\bibfield  {journal}
  {\bibinfo  {journal} {J. Chem. Phys.}\ }\textbf {\bibinfo {volume} {146}},\
  \bibinfo {pages} {084703} (\bibinfo {year} {2017})}\BibitemShut {NoStop}%
\bibitem [{\citenamefont {Sega}\ and\ \citenamefont
  {Dellago}(2017)}]{marcello2017LongRange}%
  \BibitemOpen
  \bibfield  {author} {\bibinfo {author} {\bibfnamefont {M.}~\bibnamefont
  {Sega}}\ and\ \bibinfo {author} {\bibfnamefont {C.}~\bibnamefont {Dellago}},\
  }\href {\doibase 10.1021/acs.jpcb.6b12437} {\bibfield  {journal} {\bibinfo
  {journal} {J. Phys. Chem. B}\ }\textbf {\bibinfo {volume} {121}},\ \bibinfo
  {pages} {3798} (\bibinfo {year} {2017})}\BibitemShut {NoStop}%
\bibitem [{\citenamefont {Sun}(1998)}]{sun1998compass}%
  \BibitemOpen
  \bibfield  {author} {\bibinfo {author} {\bibfnamefont {H.}~\bibnamefont
  {Sun}},\ }\href {\doibase 10.1021/jp980939v} {\bibfield  {journal} {\bibinfo
  {journal} {J. Phys. Chem. B}\ }\textbf {\bibinfo {volume} {102}},\ \bibinfo
  {pages} {7338} (\bibinfo {year} {1998})}\BibitemShut {NoStop}%
\bibitem [{\citenamefont {Sinha}\ \emph {et~al.}(2003)\citenamefont {Sinha},
  \citenamefont {Dhir}, \citenamefont {Shi}, \citenamefont {Freund},\ and\
  \citenamefont {Darve}}]{sinha2003surface}%
  \BibitemOpen
  \bibfield  {author} {\bibinfo {author} {\bibfnamefont {S.}~\bibnamefont
  {Sinha}}, \bibinfo {author} {\bibfnamefont {V.~K.}\ \bibnamefont {Dhir}},
  \bibinfo {author} {\bibfnamefont {B.}~\bibnamefont {Shi}}, \bibinfo {author}
  {\bibfnamefont {J.~B.}\ \bibnamefont {Freund}}, \ and\ \bibinfo {author}
  {\bibfnamefont {E.}~\bibnamefont {Darve}},\ }in\ \href {\doibase
  10.1115/HT2003-47164} {\emph {\bibinfo {booktitle} {ASME 2003 Heat Transfer
  Summer Conference}}}\ (\bibinfo {organization} {American Society of
  Mechanical Engineers},\ \bibinfo {year} {2003})\ pp.\ \bibinfo {pages}
  {711--714}\BibitemShut {NoStop}%
\bibitem [{\citenamefont {Werth}, \citenamefont {Horsch},\ and\ \citenamefont
  {Hasse}(2015)}]{werth2015long}%
  \BibitemOpen
  \bibfield  {author} {\bibinfo {author} {\bibfnamefont {S.}~\bibnamefont
  {Werth}}, \bibinfo {author} {\bibfnamefont {M.}~\bibnamefont {Horsch}}, \
  and\ \bibinfo {author} {\bibfnamefont {H.}~\bibnamefont {Hasse}},\ }\href
  {\doibase 10.1080/00268976.2015.1061151} {\bibfield  {journal} {\bibinfo
  {journal} {Mol. Phys.}\ }\textbf {\bibinfo {volume} {113}},\ \bibinfo {pages}
  {3750} (\bibinfo {year} {2015})}\BibitemShut {NoStop}%
\bibitem [{\citenamefont {Molinero}\ and\ \citenamefont
  {Moore}(2009)}]{molinero_water_2009}%
  \BibitemOpen
  \bibfield  {author} {\bibinfo {author} {\bibfnamefont {V.}~\bibnamefont
  {Molinero}}\ and\ \bibinfo {author} {\bibfnamefont {E.~B.}\ \bibnamefont
  {Moore}},\ }\href {\doibase 10.1021/jp805227c} {\bibfield  {journal}
  {\bibinfo  {journal} {J. Phys. Chem. B}\ }\textbf {\bibinfo {volume} {113}},\
  \bibinfo {pages} {4008} (\bibinfo {year} {2009})}\BibitemShut {NoStop}%
\bibitem [{\citenamefont {Reinhardt}\ and\ \citenamefont
  {Doye}(2012)}]{reinhardt2012free}%
  \BibitemOpen
  \bibfield  {author} {\bibinfo {author} {\bibfnamefont {A.}~\bibnamefont
  {Reinhardt}}\ and\ \bibinfo {author} {\bibfnamefont {J.~P.}\ \bibnamefont
  {Doye}},\ }\href {\doibase 10.1063/1.3677192} {\bibfield  {journal} {\bibinfo
   {journal} {J. Chem. Phys.}\ }\textbf {\bibinfo {volume} {136}},\ \bibinfo
  {pages} {054501} (\bibinfo {year} {2012})}\BibitemShut {NoStop}%
\bibitem [{\citenamefont {Cox}\ \emph {et~al.}(2015{\natexlab{a}})\citenamefont
  {Cox}, \citenamefont {Kathmann}, \citenamefont {Slater},\ and\ \citenamefont
  {Michaelides}}]{cox2015molecular1}%
  \BibitemOpen
  \bibfield  {author} {\bibinfo {author} {\bibfnamefont {S.~J.}\ \bibnamefont
  {Cox}}, \bibinfo {author} {\bibfnamefont {S.~M.}\ \bibnamefont {Kathmann}},
  \bibinfo {author} {\bibfnamefont {B.}~\bibnamefont {Slater}}, \ and\ \bibinfo
  {author} {\bibfnamefont {A.}~\bibnamefont {Michaelides}},\ }\href {\doibase
  10.1063/1.4919714} {\bibfield  {journal} {\bibinfo  {journal} {J. Chem.
  Phys.}\ }\textbf {\bibinfo {volume} {142}},\ \bibinfo {pages} {184704}
  (\bibinfo {year} {2015}{\natexlab{a}})}\BibitemShut {NoStop}%
\bibitem [{\citenamefont {Cox}\ \emph {et~al.}(2015{\natexlab{b}})\citenamefont
  {Cox}, \citenamefont {Kathmann}, \citenamefont {Slater},\ and\ \citenamefont
  {Michaelides}}]{cox2015molecular2}%
  \BibitemOpen
  \bibfield  {author} {\bibinfo {author} {\bibfnamefont {S.~J.}\ \bibnamefont
  {Cox}}, \bibinfo {author} {\bibfnamefont {S.~M.}\ \bibnamefont {Kathmann}},
  \bibinfo {author} {\bibfnamefont {B.}~\bibnamefont {Slater}}, \ and\ \bibinfo
  {author} {\bibfnamefont {A.}~\bibnamefont {Michaelides}},\ }\href {\doibase
  10.1063/1.4919715} {\bibfield  {journal} {\bibinfo  {journal} {J. Chem.
  Phys.}\ }\textbf {\bibinfo {volume} {142}},\ \bibinfo {pages} {184705}
  (\bibinfo {year} {2015}{\natexlab{b}})}\BibitemShut {NoStop}%
\bibitem [{\citenamefont {Fitzner}\ \emph {et~al.}(2015)\citenamefont
  {Fitzner}, \citenamefont {Sosso}, \citenamefont {Cox},\ and\ \citenamefont
  {Michaelides}}]{fitzner_many_2015}%
  \BibitemOpen
  \bibfield  {author} {\bibinfo {author} {\bibfnamefont {M.}~\bibnamefont
  {Fitzner}}, \bibinfo {author} {\bibfnamefont {G.~C.}\ \bibnamefont {Sosso}},
  \bibinfo {author} {\bibfnamefont {S.~J.}\ \bibnamefont {Cox}}, \ and\
  \bibinfo {author} {\bibfnamefont {A.}~\bibnamefont {Michaelides}},\ }\href
  {\doibase 10.1021/jacs.5b08748} {\bibfield  {journal} {\bibinfo  {journal}
  {J. Am. Chem. Soc.}\ }\textbf {\bibinfo {volume} {137}},\ \bibinfo {pages}
  {13658} (\bibinfo {year} {2015})}\BibitemShut {NoStop}%
\bibitem [{\citenamefont {Heinz}\ \emph {et~al.}(2008)\citenamefont {Heinz},
  \citenamefont {Vaia}, \citenamefont {Farmer},\ and\ \citenamefont
  {Naik}}]{heinz2008accurate}%
  \BibitemOpen
  \bibfield  {author} {\bibinfo {author} {\bibfnamefont {H.}~\bibnamefont
  {Heinz}}, \bibinfo {author} {\bibfnamefont {R.}~\bibnamefont {Vaia}},
  \bibinfo {author} {\bibfnamefont {B.}~\bibnamefont {Farmer}}, \ and\ \bibinfo
  {author} {\bibfnamefont {R.}~\bibnamefont {Naik}},\ }\href {\doibase
  10.1021/jp801931d} {\bibfield  {journal} {\bibinfo  {journal} {J. Phys. Chem.
  C}\ }\textbf {\bibinfo {volume} {112}},\ \bibinfo {pages} {17281} (\bibinfo
  {year} {2008})}\BibitemShut {NoStop}%
\bibitem [{\citenamefont {Xu}\ \emph {et~al.}(2015)\citenamefont {Xu},
  \citenamefont {Gao}, \citenamefont {Wang},\ and\ \citenamefont
  {Fang}}]{xu2015nanoscale}%
  \BibitemOpen
  \bibfield  {author} {\bibinfo {author} {\bibfnamefont {Z.}~\bibnamefont
  {Xu}}, \bibinfo {author} {\bibfnamefont {Y.}~\bibnamefont {Gao}}, \bibinfo
  {author} {\bibfnamefont {C.}~\bibnamefont {Wang}}, \ and\ \bibinfo {author}
  {\bibfnamefont {H.}~\bibnamefont {Fang}},\ }\href {\doibase
  10.1021/acs.jpcc.5b04237} {\bibfield  {journal} {\bibinfo  {journal} {J.
  Phys. Chem. C}\ }\textbf {\bibinfo {volume} {119}},\ \bibinfo {pages} {20409}
  (\bibinfo {year} {2015})}\BibitemShut {NoStop}%
\bibitem [{\citenamefont {Plimpton}(1995)}]{plimpton1995fast}%
  \BibitemOpen
  \bibfield  {author} {\bibinfo {author} {\bibfnamefont {S.}~\bibnamefont
  {Plimpton}},\ }\href {\doibase 10.1006/jcph.1995.1039} {\bibfield  {journal}
  {\bibinfo  {journal} {J. Comput. Phys.}\ }\textbf {\bibinfo {volume} {117}},\
  \bibinfo {pages} {1} (\bibinfo {year} {1995})}\BibitemShut {NoStop}%
\bibitem [{\citenamefont {Martyna}, \citenamefont {Klein},\ and\ \citenamefont
  {Tuckerman}(1992)}]{martyna1992nose}%
  \BibitemOpen
  \bibfield  {author} {\bibinfo {author} {\bibfnamefont {G.~J.}\ \bibnamefont
  {Martyna}}, \bibinfo {author} {\bibfnamefont {M.~L.}\ \bibnamefont {Klein}},
  \ and\ \bibinfo {author} {\bibfnamefont {M.}~\bibnamefont {Tuckerman}},\
  }\href {\doibase 10.1063/1.463940} {\bibfield  {journal} {\bibinfo  {journal}
  {J. Chem. Phys.}\ }\textbf {\bibinfo {volume} {97}},\ \bibinfo {pages} {2635}
  (\bibinfo {year} {1992})}\BibitemShut {NoStop}%
\bibitem [{\citenamefont {Laio}\ and\ \citenamefont
  {Parrinello}(2002)}]{laio2002escaping}%
  \BibitemOpen
  \bibfield  {author} {\bibinfo {author} {\bibfnamefont {A.}~\bibnamefont
  {Laio}}\ and\ \bibinfo {author} {\bibfnamefont {M.}~\bibnamefont
  {Parrinello}},\ }\href {\doibase 10.1073/pnas.202427399} {\bibfield
  {journal} {\bibinfo  {journal} {Proc. Natl. Acad. Sci. U.S.A.}\ }\textbf
  {\bibinfo {volume} {99}},\ \bibinfo {pages} {12562} (\bibinfo {year}
  {2002})}\BibitemShut {NoStop}%
\bibitem [{\citenamefont {Barducci}, \citenamefont {Bussi},\ and\ \citenamefont
  {Parrinello}(2008)}]{barducci2008well}%
  \BibitemOpen
  \bibfield  {author} {\bibinfo {author} {\bibfnamefont {A.}~\bibnamefont
  {Barducci}}, \bibinfo {author} {\bibfnamefont {G.}~\bibnamefont {Bussi}}, \
  and\ \bibinfo {author} {\bibfnamefont {M.}~\bibnamefont {Parrinello}},\
  }\href {\doibase 10.1103/PhysRevLett.100.020603} {\bibfield  {journal}
  {\bibinfo  {journal} {Phys. Rev. Lett.}\ }\textbf {\bibinfo {volume} {100}},\
  \bibinfo {pages} {020603} (\bibinfo {year} {2008})}\BibitemShut {NoStop}%
\bibitem [{\citenamefont {Tribello}\ \emph {et~al.}(2014)\citenamefont
  {Tribello}, \citenamefont {Bonomi}, \citenamefont {Branduardi}, \citenamefont
  {Camilloni},\ and\ \citenamefont {Bussi}}]{tribello2014plumed}%
  \BibitemOpen
  \bibfield  {author} {\bibinfo {author} {\bibfnamefont {G.~A.}\ \bibnamefont
  {Tribello}}, \bibinfo {author} {\bibfnamefont {M.}~\bibnamefont {Bonomi}},
  \bibinfo {author} {\bibfnamefont {D.}~\bibnamefont {Branduardi}}, \bibinfo
  {author} {\bibfnamefont {C.}~\bibnamefont {Camilloni}}, \ and\ \bibinfo
  {author} {\bibfnamefont {G.}~\bibnamefont {Bussi}},\ }\href {\doibase
  10.1016/j.cpc.2013.09.018} {\bibfield  {journal} {\bibinfo  {journal}
  {Comput. Phys. Commun.}\ }\textbf {\bibinfo {volume} {185}},\ \bibinfo
  {pages} {604} (\bibinfo {year} {2014})}\BibitemShut {NoStop}%
\bibitem [{\citenamefont {Sosso}\ \emph
  {et~al.}(2016{\natexlab{a}})\citenamefont {Sosso}, \citenamefont {Chen},
  \citenamefont {Cox}, \citenamefont {Fitzner}, \citenamefont {Pedevilla},
  \citenamefont {Zen},\ and\ \citenamefont {Michaelides}}]{sosso2016crystal}%
  \BibitemOpen
  \bibfield  {author} {\bibinfo {author} {\bibfnamefont {G.~C.}\ \bibnamefont
  {Sosso}}, \bibinfo {author} {\bibfnamefont {J.}~\bibnamefont {Chen}},
  \bibinfo {author} {\bibfnamefont {S.~J.}\ \bibnamefont {Cox}}, \bibinfo
  {author} {\bibfnamefont {M.}~\bibnamefont {Fitzner}}, \bibinfo {author}
  {\bibfnamefont {P.}~\bibnamefont {Pedevilla}}, \bibinfo {author}
  {\bibfnamefont {A.}~\bibnamefont {Zen}}, \ and\ \bibinfo {author}
  {\bibfnamefont {A.}~\bibnamefont {Michaelides}},\ }\href {\doibase
  10.1021/acs.chemrev.5b00744} {\bibfield  {journal} {\bibinfo  {journal}
  {Chem. Rev.}\ }\textbf {\bibinfo {volume} {116}},\ \bibinfo {pages} {7078}
  (\bibinfo {year} {2016}{\natexlab{a}})}\BibitemShut {NoStop}%
\bibitem [{\citenamefont {Sosso}\ \emph
  {et~al.}(2016{\natexlab{b}})\citenamefont {Sosso}, \citenamefont {Tribello},
  \citenamefont {Zen}, \citenamefont {Pedevilla},\ and\ \citenamefont
  {Michaelides}}]{sosso2016ice}%
  \BibitemOpen
  \bibfield  {author} {\bibinfo {author} {\bibfnamefont {G.~C.}\ \bibnamefont
  {Sosso}}, \bibinfo {author} {\bibfnamefont {G.~A.}\ \bibnamefont {Tribello}},
  \bibinfo {author} {\bibfnamefont {A.}~\bibnamefont {Zen}}, \bibinfo {author}
  {\bibfnamefont {P.}~\bibnamefont {Pedevilla}}, \ and\ \bibinfo {author}
  {\bibfnamefont {A.}~\bibnamefont {Michaelides}},\ }\href {\doibase
  10.1063/1.4968796} {\bibfield  {journal} {\bibinfo  {journal} {J. Chem.
  Phys.}\ }\textbf {\bibinfo {volume} {145}},\ \bibinfo {pages} {211927}
  (\bibinfo {year} {2016}{\natexlab{b}})}\BibitemShut {NoStop}%
\bibitem [{\citenamefont {Tribello}\ \emph {et~al.}(2017)\citenamefont
  {Tribello}, \citenamefont {Giberti}, \citenamefont {Sosso}, \citenamefont
  {Salvalaglio},\ and\ \citenamefont {Parrinello}}]{tribello2017analyzing}%
  \BibitemOpen
  \bibfield  {author} {\bibinfo {author} {\bibfnamefont {G.~A.}\ \bibnamefont
  {Tribello}}, \bibinfo {author} {\bibfnamefont {F.}~\bibnamefont {Giberti}},
  \bibinfo {author} {\bibfnamefont {G.~C.}\ \bibnamefont {Sosso}}, \bibinfo
  {author} {\bibfnamefont {M.}~\bibnamefont {Salvalaglio}}, \ and\ \bibinfo
  {author} {\bibfnamefont {M.}~\bibnamefont {Parrinello}},\ }\href {\doibase
  10.1021/acs.jctc.6b01073} {\bibfield  {journal} {\bibinfo  {journal} {J.
  Chem. Theory Comput.}\ }\textbf {\bibinfo {volume} {13}},\ \bibinfo {pages}
  {1317} (\bibinfo {year} {2017})}\BibitemShut {NoStop}%
\bibitem [{\citenamefont {Cheng}\ and\ \citenamefont
  {Ceriotti}(2017)}]{cheng2017bridging}%
  \BibitemOpen
  \bibfield  {author} {\bibinfo {author} {\bibfnamefont {B.}~\bibnamefont
  {Cheng}}\ and\ \bibinfo {author} {\bibfnamefont {M.}~\bibnamefont
  {Ceriotti}},\ }\href {\doibase 10.1063/1.4973883} {\bibfield  {journal}
  {\bibinfo  {journal} {J. Chem. Phys.}\ }\textbf {\bibinfo {volume} {146}},\
  \bibinfo {pages} {034106} (\bibinfo {year} {2017})}\BibitemShut {NoStop}%
\bibitem [{\citenamefont {Bussi}\ \emph {et~al.}(2006)\citenamefont {Bussi},
  \citenamefont {Gervasio}, \citenamefont {Laio},\ and\ \citenamefont
  {Parrinello}}]{bussi2006free}%
  \BibitemOpen
  \bibfield  {author} {\bibinfo {author} {\bibfnamefont {G.}~\bibnamefont
  {Bussi}}, \bibinfo {author} {\bibfnamefont {F.~L.}\ \bibnamefont {Gervasio}},
  \bibinfo {author} {\bibfnamefont {A.}~\bibnamefont {Laio}}, \ and\ \bibinfo
  {author} {\bibfnamefont {M.}~\bibnamefont {Parrinello}},\ }\href {\doibase
  10.1021/ja062463w} {\bibfield  {journal} {\bibinfo  {journal} {J. Am. Chem.
  Soc.}\ }\textbf {\bibinfo {volume} {128}},\ \bibinfo {pages} {13435}
  (\bibinfo {year} {2006})}\BibitemShut {NoStop}%
\bibitem [{\citenamefont {Laio}\ and\ \citenamefont
  {Gervasio}(2008)}]{laio2008metadynamics}%
  \BibitemOpen
  \bibfield  {author} {\bibinfo {author} {\bibfnamefont {A.}~\bibnamefont
  {Laio}}\ and\ \bibinfo {author} {\bibfnamefont {F.~L.}\ \bibnamefont
  {Gervasio}},\ }\href {\doibase 10.1088/0034-4885/71/12/126601} {\bibfield
  {journal} {\bibinfo  {journal} {Rep. Prog. Phys.}\ }\textbf {\bibinfo
  {volume} {71}},\ \bibinfo {pages} {126601} (\bibinfo {year}
  {2008})}\BibitemShut {NoStop}%
\bibitem [{\citenamefont {Bonn}\ and\ \citenamefont
  {Ross}(2001)}]{bonn2001wetting}%
  \BibitemOpen
  \bibfield  {author} {\bibinfo {author} {\bibfnamefont {D.}~\bibnamefont
  {Bonn}}\ and\ \bibinfo {author} {\bibfnamefont {D.}~\bibnamefont {Ross}},\
  }\href {\doibase 10.1088/0034-4885/64/9/202} {\bibfield  {journal} {\bibinfo
  {journal} {Rep. Prog. Phys.}\ }\textbf {\bibinfo {volume} {64}},\ \bibinfo
  {pages} {1085} (\bibinfo {year} {2001})}\BibitemShut {NoStop}%
\bibitem [{\citenamefont {Friedman}, \citenamefont {Khalil},\ and\
  \citenamefont {Taborek}(2013)}]{friedman2013wetting}%
  \BibitemOpen
  \bibfield  {author} {\bibinfo {author} {\bibfnamefont {S.~R.}\ \bibnamefont
  {Friedman}}, \bibinfo {author} {\bibfnamefont {M.}~\bibnamefont {Khalil}}, \
  and\ \bibinfo {author} {\bibfnamefont {P.}~\bibnamefont {Taborek}},\ }\href
  {\doibase doi.org/10.1103/PhysRevLett.111.226101} {\bibfield  {journal}
  {\bibinfo  {journal} {Phys. Rev. Lett.}\ }\textbf {\bibinfo {volume} {111}},\
  \bibinfo {pages} {226101} (\bibinfo {year} {2013})}\BibitemShut {NoStop}%
\bibitem [{\citenamefont {Evans}, \citenamefont {Stewart},\ and\ \citenamefont
  {Wilding}(2016)}]{evans2016critical}%
  \BibitemOpen
  \bibfield  {author} {\bibinfo {author} {\bibfnamefont {R.}~\bibnamefont
  {Evans}}, \bibinfo {author} {\bibfnamefont {M.~C.}\ \bibnamefont {Stewart}},
  \ and\ \bibinfo {author} {\bibfnamefont {N.~B.}\ \bibnamefont {Wilding}},\
  }\href {\doibase 10.1103/PhysRevLett.117.176102} {\bibfield  {journal}
  {\bibinfo  {journal} {Phys. Rev. Lett.}\ }\textbf {\bibinfo {volume} {117}},\
  \bibinfo {pages} {176102} (\bibinfo {year} {2016})}\BibitemShut {NoStop}%
\bibitem [{\citenamefont {Caupin}, \citenamefont {Sasaki},\ and\ \citenamefont
  {Balibar}(2008)}]{caupin2008absence}%
  \BibitemOpen
  \bibfield  {author} {\bibinfo {author} {\bibfnamefont {F.}~\bibnamefont
  {Caupin}}, \bibinfo {author} {\bibfnamefont {S.}~\bibnamefont {Sasaki}}, \
  and\ \bibinfo {author} {\bibfnamefont {S.}~\bibnamefont {Balibar}},\ }\href
  {\doibase 10.1088/0953-8984/20/49/494228} {\bibfield  {journal} {\bibinfo
  {journal} {J. Phys. Condens. Matter}\ }\textbf {\bibinfo {volume} {20}},\
  \bibinfo {pages} {494228} (\bibinfo {year} {2008})}\BibitemShut {NoStop}%
\bibitem [{\citenamefont {Toxvaerd}\ and\ \citenamefont
  {Dyre}(2011)}]{toxvaerd2011communication}%
  \BibitemOpen
  \bibfield  {author} {\bibinfo {author} {\bibfnamefont {S.}~\bibnamefont
  {Toxvaerd}}\ and\ \bibinfo {author} {\bibfnamefont {J.~C.}\ \bibnamefont
  {Dyre}},\ }\href {\doibase 10.1063/1.3558787} {\bibfield  {journal} {\bibinfo
   {journal} {J. Chem. Phys.}\ }\textbf {\bibinfo {volume} {134}},\ \bibinfo
  {pages} {081102} (\bibinfo {year} {2011})}\BibitemShut {NoStop}%
\bibitem [{\citenamefont {in’t Veld}, \citenamefont {Ismail},\ and\
  \citenamefont {Grest}(2007)}]{in2007application}%
  \BibitemOpen
  \bibfield  {author} {\bibinfo {author} {\bibfnamefont {P.~J.}\ \bibnamefont
  {in’t Veld}}, \bibinfo {author} {\bibfnamefont {A.~E.}\ \bibnamefont
  {Ismail}}, \ and\ \bibinfo {author} {\bibfnamefont {G.~S.}\ \bibnamefont
  {Grest}},\ }\href {\doibase 10.1063/1.2770730} {\bibfield  {journal}
  {\bibinfo  {journal} {J. Chem. Phys.}\ }\textbf {\bibinfo {volume} {127}},\
  \bibinfo {pages} {144711} (\bibinfo {year} {2007})}\BibitemShut {NoStop}%
\bibitem [{\citenamefont {Isele-Holder}, \citenamefont {Mitchell},\ and\
  \citenamefont {Ismail}(2012)}]{isele2012development}%
  \BibitemOpen
  \bibfield  {author} {\bibinfo {author} {\bibfnamefont {R.~E.}\ \bibnamefont
  {Isele-Holder}}, \bibinfo {author} {\bibfnamefont {W.}~\bibnamefont
  {Mitchell}}, \ and\ \bibinfo {author} {\bibfnamefont {A.~E.}\ \bibnamefont
  {Ismail}},\ }\href {\doibase 10.1063/1.4764089} {\bibfield  {journal}
  {\bibinfo  {journal} {J. Chem. Phys.}\ }\textbf {\bibinfo {volume} {137}},\
  \bibinfo {pages} {174107} (\bibinfo {year} {2012})}\BibitemShut {NoStop}%
\bibitem [{\citenamefont {Isele-Holder}\ \emph {et~al.}(2013)\citenamefont
  {Isele-Holder}, \citenamefont {Mitchell}, \citenamefont {Hammond},
  \citenamefont {Kohlmeyer},\ and\ \citenamefont
  {Ismail}}]{isele2013reconsidering}%
  \BibitemOpen
  \bibfield  {author} {\bibinfo {author} {\bibfnamefont {R.~E.}\ \bibnamefont
  {Isele-Holder}}, \bibinfo {author} {\bibfnamefont {W.}~\bibnamefont
  {Mitchell}}, \bibinfo {author} {\bibfnamefont {J.~R.}\ \bibnamefont
  {Hammond}}, \bibinfo {author} {\bibfnamefont {A.}~\bibnamefont {Kohlmeyer}},
  \ and\ \bibinfo {author} {\bibfnamefont {A.~E.}\ \bibnamefont {Ismail}},\
  }\href {\doibase 10.1021/ct4004614} {\bibfield  {journal} {\bibinfo
  {journal} {J. Chem. Theory Comput.}\ }\textbf {\bibinfo {volume} {9}},\
  \bibinfo {pages} {5412} (\bibinfo {year} {2013})}\BibitemShut {NoStop}%
\bibitem [{\citenamefont {Joly}(2011)}]{joly2011capillary}%
  \BibitemOpen
  \bibfield  {author} {\bibinfo {author} {\bibfnamefont {L.}~\bibnamefont
  {Joly}},\ }\href {\doibase 10.1063/1.3664622} {\bibfield  {journal} {\bibinfo
   {journal} {J. Chem. Phys.}\ }\textbf {\bibinfo {volume} {135}},\ \bibinfo
  {pages} {214705} (\bibinfo {year} {2011})}\BibitemShut {NoStop}%
\bibitem [{\citenamefont {Gravelle}\ \emph {et~al.}(2016)\citenamefont
  {Gravelle}, \citenamefont {Ybert}, \citenamefont {Bocquet},\ and\
  \citenamefont {Joly}}]{gravelle2016anomalous}%
  \BibitemOpen
  \bibfield  {author} {\bibinfo {author} {\bibfnamefont {S.}~\bibnamefont
  {Gravelle}}, \bibinfo {author} {\bibfnamefont {C.}~\bibnamefont {Ybert}},
  \bibinfo {author} {\bibfnamefont {L.}~\bibnamefont {Bocquet}}, \ and\
  \bibinfo {author} {\bibfnamefont {L.}~\bibnamefont {Joly}},\ }\href {\doibase
  10.1103/PhysRevE.93.033123} {\bibfield  {journal} {\bibinfo  {journal} {Phys.
  Rev. E}\ }\textbf {\bibinfo {volume} {93}},\ \bibinfo {pages} {033123}
  (\bibinfo {year} {2016})}\BibitemShut {NoStop}%
\bibitem [{\citenamefont {Hens}, \citenamefont {Agarwal},\ and\ \citenamefont
  {Biswas}(2014)}]{hens2014nanoscale}%
  \BibitemOpen
  \bibfield  {author} {\bibinfo {author} {\bibfnamefont {A.}~\bibnamefont
  {Hens}}, \bibinfo {author} {\bibfnamefont {R.}~\bibnamefont {Agarwal}}, \
  and\ \bibinfo {author} {\bibfnamefont {G.}~\bibnamefont {Biswas}},\
  }\href@noop {} {\bibfield  {journal} {\bibinfo  {journal} {Int. J. Heat Mass
  Transfer}\ }\textbf {\bibinfo {volume} {71}},\ \bibinfo {pages} {303}
  (\bibinfo {year} {2014})}\BibitemShut {NoStop}%
\bibitem [{\citenamefont {Nagayama}\ \emph {et~al.}(2015)\citenamefont
  {Nagayama}, \citenamefont {Takematsu}, \citenamefont {Mizuguchi},\ and\
  \citenamefont {Tsuruta}}]{nagayama2015molecular}%
  \BibitemOpen
  \bibfield  {author} {\bibinfo {author} {\bibfnamefont {G.}~\bibnamefont
  {Nagayama}}, \bibinfo {author} {\bibfnamefont {M.}~\bibnamefont {Takematsu}},
  \bibinfo {author} {\bibfnamefont {H.}~\bibnamefont {Mizuguchi}}, \ and\
  \bibinfo {author} {\bibfnamefont {T.}~\bibnamefont {Tsuruta}},\ }\href
  {\doibase 10.1063/1.4923261} {\bibfield  {journal} {\bibinfo  {journal} {J.
  Chem. Phys.}\ }\textbf {\bibinfo {volume} {143}},\ \bibinfo {pages} {014706}
  (\bibinfo {year} {2015})}\BibitemShut {NoStop}%
\bibitem [{\citenamefont {Iyer}, \citenamefont {Mendenhall},\ and\
  \citenamefont {Blankschtein}(2013)}]{iyer2013computer}%
  \BibitemOpen
  \bibfield  {author} {\bibinfo {author} {\bibfnamefont {J.}~\bibnamefont
  {Iyer}}, \bibinfo {author} {\bibfnamefont {J.~D.}\ \bibnamefont
  {Mendenhall}}, \ and\ \bibinfo {author} {\bibfnamefont {D.}~\bibnamefont
  {Blankschtein}},\ }\href {\doibase 10.1021/jp4001253} {\bibfield  {journal}
  {\bibinfo  {journal} {J. Phys. Chem. B}\ }\textbf {\bibinfo {volume} {117}},\
  \bibinfo {pages} {6430} (\bibinfo {year} {2013})}\BibitemShut {NoStop}%
\bibitem [{\citenamefont {Tran}\ and\ \citenamefont
  {Kalra}(2014)}]{tran2014molecular}%
  \BibitemOpen
  \bibfield  {author} {\bibinfo {author} {\bibfnamefont {C.}~\bibnamefont
  {Tran}}\ and\ \bibinfo {author} {\bibfnamefont {V.}~\bibnamefont {Kalra}},\
  }\href {\doibase 10.1063/1.4869404} {\bibfield  {journal} {\bibinfo
  {journal} {J. Chem. Phys.}\ }\textbf {\bibinfo {volume} {140}},\ \bibinfo
  {pages} {134902} (\bibinfo {year} {2014})}\BibitemShut {NoStop}%
\bibitem [{\citenamefont {Radu}\ and\ \citenamefont
  {Kremer}(2017)}]{radu2017enhanced}%
  \BibitemOpen
  \bibfield  {author} {\bibinfo {author} {\bibfnamefont {M.}~\bibnamefont
  {Radu}}\ and\ \bibinfo {author} {\bibfnamefont {K.}~\bibnamefont {Kremer}},\
  }\href {\doibase 10.1103/PhysRevLett.118.055702} {\bibfield  {journal}
  {\bibinfo  {journal} {Phys. Rev. Lett.}\ }\textbf {\bibinfo {volume} {118}},\
  \bibinfo {pages} {055702} (\bibinfo {year} {2017})}\BibitemShut {NoStop}%
\bibitem [{\citenamefont {Reinhardt}\ and\ \citenamefont
  {Doye}(2014)}]{reinhardt2014effects}%
  \BibitemOpen
  \bibfield  {author} {\bibinfo {author} {\bibfnamefont {A.}~\bibnamefont
  {Reinhardt}}\ and\ \bibinfo {author} {\bibfnamefont {J.~P.}\ \bibnamefont
  {Doye}},\ }\href {\doibase 10.1063/1.4892804} {\bibfield  {journal} {\bibinfo
   {journal} {J. Chem. Phys.}\ }\textbf {\bibinfo {volume} {141}},\ \bibinfo
  {pages} {084501} (\bibinfo {year} {2014})}\BibitemShut {NoStop}%
\bibitem [{\citenamefont {Cabriolu}\ and\ \citenamefont
  {Li}(2015)}]{cabriolu2015ice}%
  \BibitemOpen
  \bibfield  {author} {\bibinfo {author} {\bibfnamefont {R.}~\bibnamefont
  {Cabriolu}}\ and\ \bibinfo {author} {\bibfnamefont {T.}~\bibnamefont {Li}},\
  }\href {\doibase 10.1103/PhysRevE.91.052402} {\bibfield  {journal} {\bibinfo
  {journal} {Phys. Rev. E}\ }\textbf {\bibinfo {volume} {91}},\ \bibinfo
  {pages} {052402} (\bibinfo {year} {2015})}\BibitemShut {NoStop}%
\bibitem [{\citenamefont {Bi}, \citenamefont {Cabriolu},\ and\ \citenamefont
  {Li}(2016)}]{bi2016heterogeneous}%
  \BibitemOpen
  \bibfield  {author} {\bibinfo {author} {\bibfnamefont {Y.}~\bibnamefont
  {Bi}}, \bibinfo {author} {\bibfnamefont {R.}~\bibnamefont {Cabriolu}}, \ and\
  \bibinfo {author} {\bibfnamefont {T.}~\bibnamefont {Li}},\ }\href {\doibase
  10.1021/acs.jpcc.5b09740} {\bibfield  {journal} {\bibinfo  {journal} {J.
  Phys. Chem. C}\ }\textbf {\bibinfo {volume} {120}},\ \bibinfo {pages} {1507}
  (\bibinfo {year} {2016})}\BibitemShut {NoStop}%
\bibitem [{\citenamefont {Qiu}\ and\ \citenamefont
  {Molinero}(2017)}]{qiu2017strength}%
  \BibitemOpen
  \bibfield  {author} {\bibinfo {author} {\bibfnamefont {Y.}~\bibnamefont
  {Qiu}}\ and\ \bibinfo {author} {\bibfnamefont {V.}~\bibnamefont {Molinero}},\
  }\href {\doibase 10.3390/cryst7030086} {\bibfield  {journal} {\bibinfo
  {journal} {Crystals}\ }\textbf {\bibinfo {volume} {7}},\ \bibinfo {pages}
  {86} (\bibinfo {year} {2017})}\BibitemShut {NoStop}%
\bibitem [{\citenamefont {Bourque}, \citenamefont {Locker},\ and\ \citenamefont
  {Rutledge}(2017)}]{bourque2017heterogeneous}%
  \BibitemOpen
  \bibfield  {author} {\bibinfo {author} {\bibfnamefont {A.~J.}\ \bibnamefont
  {Bourque}}, \bibinfo {author} {\bibfnamefont {C.~R.}\ \bibnamefont {Locker}},
  \ and\ \bibinfo {author} {\bibfnamefont {G.~C.}\ \bibnamefont {Rutledge}},\
  }\href {\doibase 10.1021/acs.jpcb.6b12590} {\bibfield  {journal} {\bibinfo
  {journal} {J. Phys. Chem. B}\ }\textbf {\bibinfo {volume} {121}},\ \bibinfo
  {pages} {904} (\bibinfo {year} {2017})}\BibitemShut {NoStop}%
\end{thebibliography}%

\end{document}